\pdfoutput=1 %arXiv admin added this line
\documentclass[aps,prl,twocolumn,superscriptaddress]{revtex4}

\usepackage{graphicx}
\usepackage{color}

\begin{document}

\title{{\it Ab-initio} Dynamics of  Rare Thermally Activated  Reactions}

\author{S. a Beccara}
\affiliation{Physics Department of Trento University, Via Sommarive 14, Povo (Trento), I-38100, Italy.}

\author{G. Garberoglio}
\affiliation{CNISM and Physics Department of Trento University, Via Sommarive 14, Povo (Trento), I-38100, Italy.}

\author{P. Faccioli}
\email[Corresponding author: ]{faccioli@science.unitn.it}
\affiliation{Physics Department of Trento University and INFN, Via Sommarive 14, Povo (Trento), I-38100, Italy.}
\author{F. Pederiva}
\affiliation{Physics Department of Trento University and INFN, Via Sommarive 14, Povo (Trento), I-38100, Italy.}

\begin{abstract}
We introduce a framework to investigate {\it ab-initio} the dynamics of rare thermally activated reactions. The electronic degrees of freedom are described at the quantum-mechanical level in the Born-Oppenheimer approximation, while the nuclear degrees of freedom are coupled to a thermal bath, through a Langevin equation. This method is based on the path integral representation for the stochastic dynamics  and yields the time evolution of both nuclear and electronic 
degrees of freedom, along the most probable reaction pathways, without spending computational time to explore metastable states. 
This approach is very  efficient and allows to study  thermally activated reactions which cannot be simulated using \textit{ab-initio} molecular dynamics techniques.
As a first illustrative application, we characterize the dominant pathway in  the cyclobutene$\rightarrow$butadiene reaction. 
\end{abstract}

\pacs{}

\maketitle

Thermally activated conformational and chemical reactions drive a large class of phenomena  in  physics, chemistry and biology. 
From the  theoretical standpoint, the understanding of such processes involves the characterization of the ensemble of statistically significant reaction pathways, i.e. the set of consecutive system's configurations which are most likely to be visited
during the transition from the reactant to the product state, at a given finite temperature. 
Such pathways provide information about the structure of the transition state.%, and therefore allow to make contact with kinetics experiments. 

In this  context,  the study of the dynamics of the reaction within an {\it ab-initio} approach is particularly  challenging. In fact, a formalism should be adopted which combines the quantum description of the electronic degrees of freedom with 
 the out-of-equilibrium dynamics generated by
the coupling of the nuclear degrees of freedom with the heat bath. {\it Ab-initio} Molecular Dynamics (MD) simulations   satisfy this requirement, at least in principle.  
In practice, all MD methods become very inefficient as soon as the activation barriers are of the order of few $k_B T$ or larger. 
The reason is that most of the computational time is spent to explore meta-stable states.

In view of such difficulties, a variety of different methods have been proposed, which neglect the dynamics arising from the coupling with the heat-bath, but provide information about the structure of the potential energy surface in the transition 
region. 
%One of the most popular approaches among the theoretical chemists community is based on the location of first-order saddle points of the molecular potential energy\cite{firstorder}. 
%For small molecules a widespread method is  the eigenvector-following algorithm, initially proposed by Baker \cite{baker}.
%In large molecular systems, though, such an approach rapidly becomes prohibitively heavy. As an alternative, 
In particular, methods based on the search of the minimum energy pathways  have been proposed, 
such as the Nudged Elastic Band (NEB) ~\cite{NEB}. Such approaches are based on the minimization of the total energy of a virtual  chain of molecular configurations connecting the reactant and product states. They provide an estimate of the activation barriers, but do not account for the dynamics and thermodynamics of the reaction. For example, there is no guarantee that a NEB
path is representative of the ensemble of statically significant reaction paths at a given temperature.  
%Methods such as biased-dynamics~\cite{biasedmd},  hyper-dynamics\cite{hypermd}  allow to reconstruct the structure of the free-energy surface. 
%However, they do not apply to out-of-equilibrium dynamics.  
Methods such as the temperature-accelerated MD \cite{tempMD} allow in principle to assess long-time dynamics, but they only apply to first-order reactions, are based on the approximations of the transition state theory, and are mostly efficient for small barrier problems.

The Dominant Reaction Pathway approach (DRP)\cite{DRP1, DRP2, DRP3, DRP4} is a recently developed rigorous framework to identify the most probable transition paths in systems described by the over-damped Langevin dynamics.  
%The method is based on the path integral representation of the  solution of the associated Fokker-Planck equation and takes into account the coupling with a heat-bath, both  in and out of thermal equilibrium.  
The DRP approach is particularly efficient  when applied to study thermally activated transitions,  since it does not waste computational time to explore metastable states. 
In addition, it directly yields  the most probable paths without the need to perform a statistical analysis of a representative sample of reactive trajectories. It also offers a natural definition of the reaction coordinate, parametrized in terms of the total Euclidean distance covered by all atoms along the dominant trajectory. Finally, the method yields information about the dynamics, since it provides the most probable time evolution of the system's degrees of freedom,  
during the reaction.

So far, the DRP approach has been applied to study conformational reactions, in particular protein folding~\cite{DRP1, DRP4}, using empirical force fields. 
In this Letter, we extend  this method and present its first application to an {\it ab-initio} simulation of a rare transition.  This is done by  evaluating the molecular energy which enters the DRP equations directly from quantum-mechanical calculations. 
As a result the DRP method yields the most probable nuclear trajectories and the corresponding electron densities, in the Born-Oppenheimer approximation. This improvement opens the door to a systematic
exploration of the dynamics of the electron density in reactions characterized by large activation energies.

Let us begin by briefly reviewing the DRP approach.  The over-damped Langevin equation describing a system of  $N$ atoms of nuclear coordinates ${\bf X}(t)\equiv ({\bf x}_1(t),\ldots, {\bf x}_N(t))$,  in contact with a thermal-bath at temperature $T$ reads:
\begin{equation} \label{langev_eq}
 \frac{\partial {\bf X} } {\partial t} = -\frac{D}{k_B T} {\bf \nabla}U({\bf X}) + {\bf \eta}(t)
\end{equation} 
where $D$ is the diffusion coefficient, $U({\bf X})$ is the molecular potential energy, and ${\bf \eta}(t)$ is a random noise with Gaussian distribution, zero average and correlation given by $\langle \eta_i(t) \eta_j(t') \rangle = 2D \ \delta_{ij} \ \delta(t-t')$.

The conditional probability density for the system to be found in the configuration ${\bf X}_f$ at time $t_f$, provided it was prepared in some initial configuration ${\bf X}_i$ at time $t_i$ is the Green's function of the Fokker-Planck operator 
associated to Eq. (\ref{langev_eq}) and can 
be represented in the following path integral form:
\begin{eqnarray} \label{probab_path}
 P({\bf X}_f, t_f | {\bf X}_i, t_i ) = e^{-\frac{U({\bf X}_f) - U( {\bf X}_i ) }{2k_BT} }
\int_{{\bf X}_i}^{{\bf X}_f} \mathcal{D} {\bf X}(t) e^{-S_{eff} [ \mathbf{X}]},
\end{eqnarray} 
where $S_{eff}$ is called the effective action, and is given by
\begin{eqnarray} \label{effact}
 S_{eff}[{X}] = \int_{t_i}^{t_f}d\tau~\frac {\dot{\bf X}^2(\tau)} {4D} + V_{eff} \left( {\bf X}(\tau) \right).
\end{eqnarray}
$V_{eff} ({\bf X})$ is called the effective potential, and reads
\begin{equation} \label{eff_pot}
 V_{eff} ({\bf X}) = \frac{D}{4(k_B T)^2 } \left[ \left| {\bf \nabla}U({\bf X})  \right| ^2 - 2k_B
T\; \nabla^2U({\bf X}) \ \right].
\end{equation} 

If ${\bf X}_f$ and ${\bf X}_i$ are product and reactant configurations, respectively, then the factor $\exp(- S_{eff}[{\bf X}])$ inside the path integral expression (\ref{probab_path}) represents the statistical weight of a given 
reactive trajectory ${\bf X}(t)$. 
In particular, the most probable trajectories are those which minimize the effective action functional $S_{eff}[{\bf X}]$. 
These are  the solutions of the classical equations of motion generated by the effective Lagrangian $\mathcal{L} = \frac { \dot { {\mathbf X}}^2(t)} {4D}  + V_{eff} \left( {\mathbf X}(t) \right) $, with boundary condition ${\bf X}(t_i)={\bf X}_i$ and ${\bf X}(t_f)={\bf X}_f$.

%\begin{figure}[t]
%\includegraphics[width=8cm]{reaction.pdf}
%\caption{The cyclobutene-butadiene reaction investigated with the ab-initio DRP approach.}
%\label{fig1}
%\end{figure}

The numerical advantage of the DRP approach follows from observing that such equations of motion conserve the effective energy $E_{eff}= \frac { \dot { {\mathbf X}}^2(t)} {4D}  - V_{eff} \left( {\mathbf X}(t) \right). $ 
This property allows  to switch from the {\it time}-dependent Newtonian description to the {\it energy}-dependent Hamilton-Jacobi (HJ) description. 
In the HJ framework the most probable pathways connecting the given initial and final configurations are obtained by minimizing the target HJ~functional
%\begin{equation} \label{hj-action}
$ S_{HJ}[{\bf X} ]= \int_{{\bf X}_i}^{{\bf X}_f} dl \; \sqrt{ \frac{1}{D} \left[ E_{eff} + V_{eff}\left({\bf X}(l) \right) \right]},
$%\end{equation} 
where $dl = \sqrt{d {\bf X}^2}$ is the  elementary  Euclidean distance in configuration space,  along the dominant reaction path. 

The effective energy parameter $E_{eff}$ fixes the total time $t_{tot}=t({\bf X}_f)$ for a single barrier-crossing transition.
In fact, the time $t \left[{\bf X} \right] $ at which the configuration ${\bf X}$ is visited, during the most probable reaction pathway is given by:
\begin{eqnarray}
t \left( {\bf X} \right) = \int_{{\bf X}_i}^{{\bf X}} dl \frac{1}{\sqrt{4 D \left[ E_{eff}+V_{eff}\left( {\bf X}(l) \right) \right]}}.
\label{time}
 \end{eqnarray}
In the present work,  we have chosen $E_{eff}=-V_{eff}({\bf X}_f)$ which yields the longest total time $t_{tot}$~
\cite{DRP2, DRP3}. 
 We stress the fact that  $t_{tot}$ is much shorter than the mean-first-passage time, as it corresponds to the 
time to reach the product, {\it once the system has left the reactant}.

In practice, finding the dominant reaction pathway amounts to minimizing a discretized version of the effective HJ functional:
\begin{eqnarray} \label{effact_discr}
 S_{HJ}^{d} [{\bf X}]= \sum_{i=1}^{N_s-1} \sqrt{ \frac{1}{D} \left[ E_{eff} + V_{eff}\left( {\bf X}_i \right) \right] } \; \Delta l_{i, i+1} \nonumber
\end{eqnarray}
where $N_s$ is the number of path discretization steps, and  the effective potential $V_{eff}({\bf X})$ is determined according to (\ref{eff_pot})  from the molecular potential energy $U({\bf X})$. The latter corresponds to the lowest eigenvalue  of the
Schr\"odinger equation for the electronic degrees of freedom, in which the nuclear coordinates are held fixed at the configuration ${\bf X}$. 
$\Delta l_{i, i+1}$ is the Euclidean distance between the slices $i$ and $i+1$, i.e.
$%\begin{equation}
  \Delta l_{i, i+1} = \sqrt{ \left| {\bf X}_{i+1} - {\bf X}_{i} \right| ^2 }.
$%\end{equation} 

\begin{figure}
\includegraphics[bb=0 0 792 612,width=8cm]{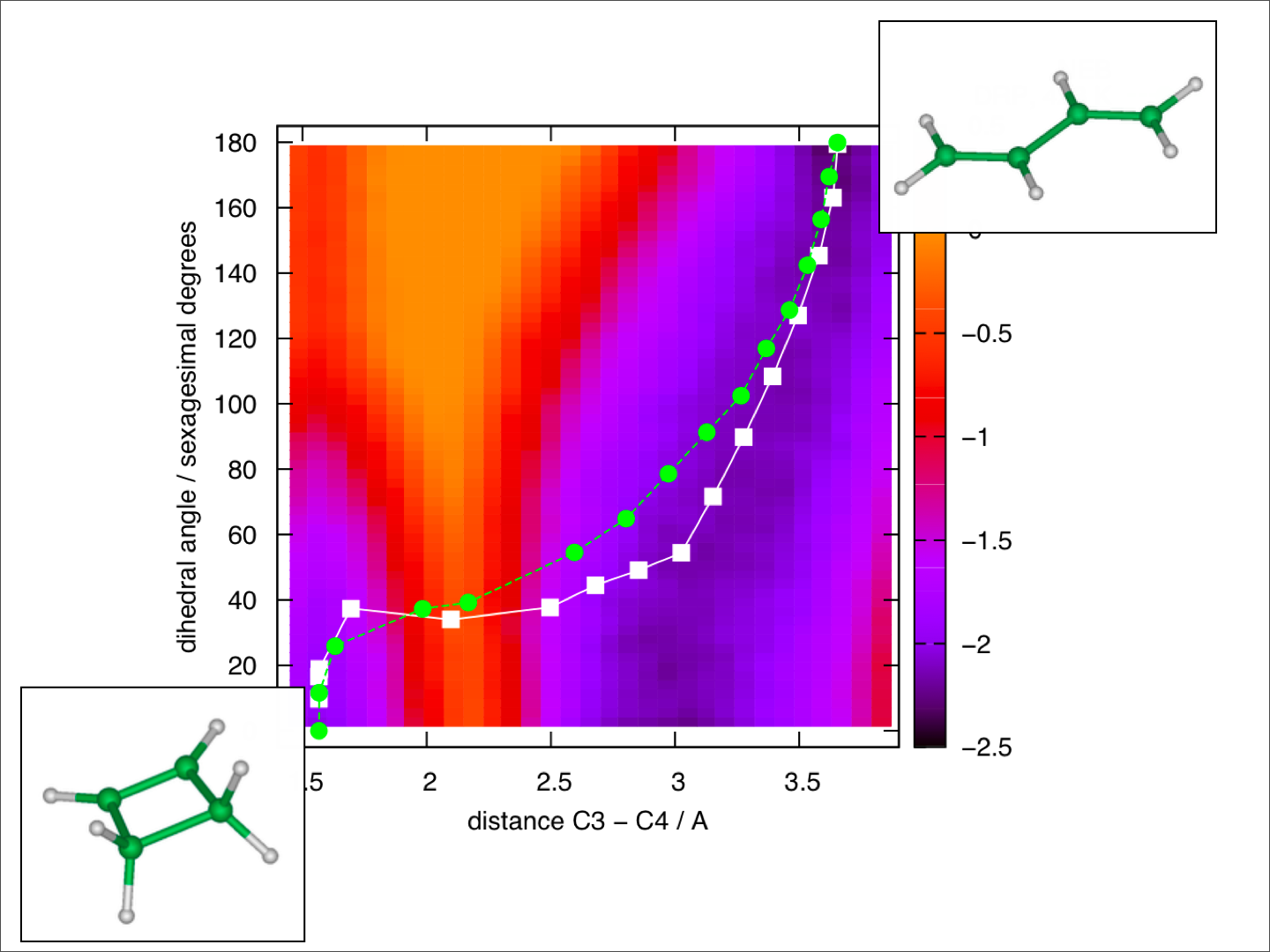}
\caption{Free-energy diagram for the reaction cyclobutene $\rightarrow$ butadiene. Superimposed the NEB pathway (squares) and DRP pathway (circles), evaluated  at $T=423 K$.}
\label{fig1}
\end{figure}
The global minimization of the HJ effective action  (\ref{effact_discr}) is a delicate task. The difficulties arise from the  ruggedness  of the effective potential, which depends both on the gradient and on the laplacian of the molecular potential $U({\bf X})$.
% Another issue is that, for methods requiring the gradient of the effective action, the derivatives of the Laplacian must be calculated. In our case, this had to be done numerically, thus introducing a source of numerical inaccuracy, and considerably slowing down the calculation.
In order to perform such minimization, we tried different approaches, including simulated annealing and first-order methods such as conjugate gradients or the Broyden-Fletcher-Goldfarb-Shanno method \cite{BFGS}. 
%We observed that the simulated annealing requires a very high fictitious temperature to significantly sample the relevant portions of configuration space, due to the very high barriers which must be overcome. 
%The convergence to the minimum is thus exceedingly slow. On the other hand, first-order algorithms are faster, but  get stuck in local minima. 
The  method which has performed best among the ones we tried is the Fast Inertial Relaxation Engine (FIRE)~\cite{FIRE}.
The noise in the numerical gradients of the HJ action becomes a limiting factor, when the gradient norm descends below a certain threshold. Therefore, we employed a simplex algorithm for the final stage of the minimization.

In the discretized representation of the HJ effective action (\ref{effact_discr}), the width of the distribution of the Euclidean distances between consecutive path slices, $\Delta l_{i,i+1}$, should not be allowed to increase in an uncontrolled way, in order to prevent all frames to 
collapse into the reactant or product configurations. A simple way to achieve this goal is  to include a penalty function based on the standard deviation of the distances~\cite{Elber}, i.e. $S^d_{HJ}\left [{\bf X} ] \rightarrow S_{HJ}[{\bf X}] + C[{\bf X} \right]$,
with
\begin{eqnarray}
 C[{\bf X}(l)] = \lambda  \left( \langle \Delta l ^2 \rangle - \langle \Delta l \rangle ^2 \right) \equiv \lambda \sigma ^2, 
\end{eqnarray} 
where $\langle {\Delta l} \rangle =\frac{1}{N_s-1}\sum_{i=1}^{N_s-1}\Delta l_{i,i+1}$, $\langle {\Delta l^2} \rangle =\frac{1}{N_s-1}\sum_{i=1}^{N_s-1}(\Delta l_{i,i+1} )^2$
and $\lambda$ is an adjustable parameter. 
The penalty function  limits the fluctuations of the  Euclidean distances $\Delta l_{i,i+1}$  and becomes irrelevant in the limit of very large number of path steps $N_s$. Unfortunately, for a typical number of
 slices $N_s\simeq 20-50$, this term was seen to introduce significant bias. 
In order to overcome this difficulty, we introduced a Lagrange multiplier in the FIRE algorithm, which holds fixed at $0.1$ the ratio between the mean-square deviation from the average of the inter-slice distances $\sigma^2$ %$\Delta l_{i,i+1}$, 
and the square of the average inter-slice distance $\langle \Delta l \rangle$.

\begin{figure}
 \includegraphics[bb=0 0 792 612,width=8 cm]{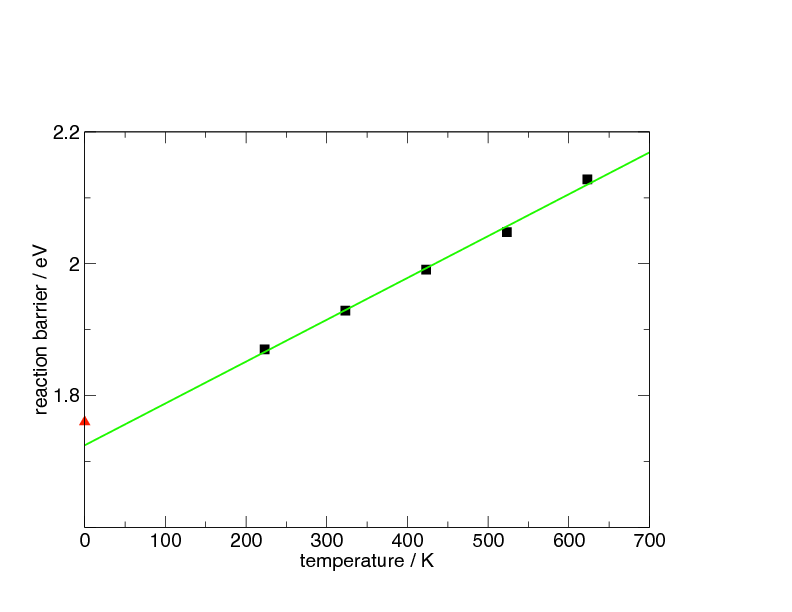}
\caption{Reaction barrier evaluated along the dominant pathways at different temperatures. The triangle represents the NEB result, within the same semi-empiric approach.}
\label{fig2}
\end{figure}

Let us now discuss the \textit{ab-initio} evaluation of the molecular energy $U({\bf X})$ and of its first and second derivatives, which enter in the definition of the effective potential (\ref{eff_pot}). 
This task requires to compute the approximate ground-state energy 
of the Schr\"odinger equation for the electronic degrees of freedom, in the Born-Oppenheimer approximation. 
In this first exploratory work, we choose to keep calculations as short as possible and adopted the PM3  semi-empirical approach in the  MOPAC implementation\cite{MOPAC}.%, which requires a modest investment of computational time.
The pathway so obtained may be used as a starting point for more accurate calculations, based, e.g., on density functional theory.

As a first  application of the {\it ab-initio} DRP  method, we computed the dominant pathway for the reaction cyclobutene $\rightarrow$ butadiene shown in 
 Fig.\ref{fig1}, at the temperature $T=423~$K. 
In the course of this reaction an energy barrier of $\sim 2~$ eV is overcome \cite{libro} and therefore this process can hardly be studied with {\it ab-initio} MD techniques. 
The DRP method was implemented using 16 discretization slices
% and the minimization algorithm was parallelized by assigning each path slice a different processor. 
and the  identification of the dominant reaction path required 80 CPU hours on 3~GHz processors.
We verified  that the FIRE algorithm achieved convergence to the same  minimum, irrespective of the starting path. 
%We tried either a linear path, or paths sampled from a Metropolis Monte Carlo at $\mathrm{1x10^6}$ K, or from a molecular dynamics run at room temperature. The relative standard deviation of the average interslice distance was kept at 10\%. 

In Fig. \ref{fig1} we present the resulting dominant reaction path, projected on  the plane defined by the distance between C3 and C4 and the dihedral angle defined by C1, C2 and the two  hydrogens bonded to them. 
The DRP path (circles) is compared with the NEB path (squares) and superimposed on the free-energy map computed  from {\it ab-initio} meta-dynamics~\cite{laio} simulations.
\begin{figure}[t!]
 \includegraphics[bb=0 0 842 595,width=8 cm]{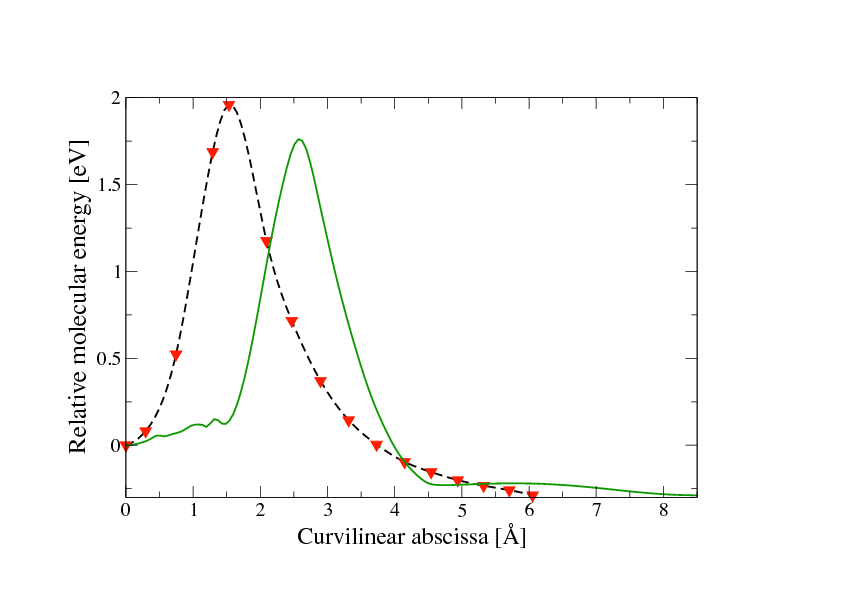}
\caption{Relative molecular energy along the DRP (dashed) and NEB (solid) paths,  parametrized in terms of the curvilinear abscissa $l$, which corresponds to the total distance covered by the atomic nuclei during the reaction.}
\label{fig3}
\end{figure}

First of all, we observe that the DRP  at this temperature and NEB paths are quite different. While the minimum energy path cuts across the iso-free-energy lines, the  most probable path  takes a shorter route, remaining for a longer portion in the high free-energy region. This can be explained by the fact that  the dominant paths are the result of a compromise between minimizing the total length of the path distance and the factor $\sqrt{E_{eff}+V_{eff}[{\bf X}(l)]}$, which appears in the integrand. 
In particular, in the high temperature limit, the free energy landscape becomes flat  and the most probable path reduces to a straight line in configuration space. 
On the other hand,  as the temperature  is reduced, the ${\bf \nabla} U({\bf X})$  contribution to the effective potential becomes more and more important, and the path should approach the NEB result.
In order to further study the temperature dependence of the DRP paths we compare in Fig. \ref{fig2} the difference between the highest molecular energy reached along the dominant path and the reactant molecular energy, for different temperatures. We observe that this quantity grows linearly with temperature. A simple linear extrapolation to zero temperature yields an energy barrier of 1.72~eV, which is close to the NEB result 
%for the relative energy difference at the saddle point 
calculated within the same semi-empirical approach, 1.76~eV.
These results show that  thermal effects can sizeably affect the structure of the statistically important paths at finite temperature. In particular, the most probable path does not in general pass through the saddle point of 
the molecular energy. 

It is also instructive to compare the value of molecular energies along the  NEB and DRP paths, reported in Fig.~\ref{fig3}.
The NEB curve displays a small secondary maximum, which is usually interpreted as a rotational  barrier for the unsaturated hydrogen~\cite{libro}. On the other hand, our DRP result does not show this feature, in the statistically most relevant path. This implies that the opening of the ring and the rotation of the unsaturated hydrogen are most likely to occur simultaneously, given the small value of the rotational barrier (about 0.01 eV) obtained at the PM3 level.
\begin{figure}
 \includegraphics[bb=0 0 842 595,width=8 cm]{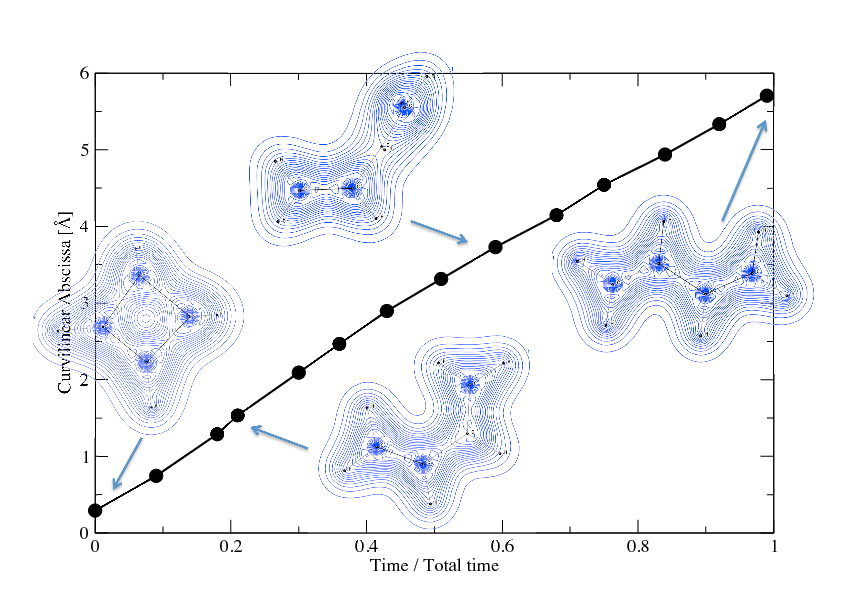}
\caption{Relative time at which each configuration is visited. Super-imposed are the  electron iso-density surface lines, on the plane selected by the C1-C3-C5
 atoms. The absolute value of the total barrier-crossing time $t_{tot}$ depends on the diffusion coefficient, and therefore on the details of the experimental conditions in which the reaction is realized.}
\label{fig4}
\end{figure}

The  {\it ab-initio} DRP approach does not only yield pathways in configuration space, but also allows to determine the time evolution 
of both nuclear and electronic degrees of freedom.
As an illustration, Fig.~\ref{fig4} shows the relative time $t(l)/t_{tot}$ at which each configuration is visited. Super-imposed are the  electron iso-density surface lines, on the plane selected by the C1-C3-C4
 atoms, at four different instants of the reaction.
 
In conclusion, in this work we have developed an {\it ab-initio} approach  to investigate the dynamics of molecular systems in contact with a heat-bath, and to find the most probable reaction pathway. Its main advantage is that its computational cost does not depend on the height of the barrier. Thus, for thermally activated reactions, the resources required are drastically smaller than that of {\it ab-initio}
MD algorithms such as  Car-Parrinello. This improvement opens the door to systematically explore the time evolution of 
the positions of the atomic nuclei and of the electron densities during reactions characterized by large activation energies, and  can be used to study the temperature dependence of the reaction mechanism.
As a first application, we have applied our method  to investigate a the cyclobutene $\rightarrow$ butadiene reaction. 
We have found that, that the coupling with the heat-bath can appreciably  affect the structure of the statistically important paths, at finite temperature.
The information which can be obtained from DRP is therefore complementary to that accessible from NEB. In fact, the latter can accurately predict the location of the saddle-point, but cannot investigate the dynamics of the statistically
significant  transitions.

\vspace{0.2 cm}

Calculations were performed on the WIGLAF cluster of the Physics Department of Trento University. 
The DRP method was developed in collaboration with M. Sega and H. Orland. We acknowledge important discussions with W.J. Lester.


\begin{thebibliography}{99}
\bibitem{NEB}D. Sheppard, R. Terrell, and G. Henkelman, J. Chem. Phys. {\bf 128}, 134106, (2008) 
\bibitem{tempMD} D. Hamelberg, J. Mongan, and J.A.  McCammon,  J. Chem. Phys. {\bf 120},  11919, (2004) 
\bibitem{DRP1}P. Faccioli, M. Sega, F. Pederiva and H. Orland,  Phys. Rev. Lett. {\bf 97}, 108101, (2006)
\bibitem{DRP2} M. Sega, P. Faccioli,  F. Pederiva, G. Garberoglio and H. Orland, Phys. Rev. Lett. {\bf 99}, 118102, (2007)  
\bibitem{DRP3} E. Autieri, P. Faccioli, M. Sega, F. Pederiva and H. Orland,   J. Chem Phys. {\bf 130}, 064106, (2009)
\bibitem{DRP4} P. Faccioli,   J.  Phys. Chem.  B {\bf 112} (2008) 13756.
\bibitem{BFGS} J. Nocedal, S.J.  Wright,  "Numerical Optimization" (2nd ed.), Springer-Verlag, Berlin, New York, 2006.
\bibitem{FIRE} E. Bitzek, P. Koskinen, F. G\"a hler, M. Moseler and P. Gumbsch, Phys. Rev. Lett.  {\bf 97}, 170201, (2006)
 \bibitem{Elber} A. Ghosh, R. Elber and H.A. Sheraga, Proc. Nat. Acad. Sci. {\bf 99}, 10394 (2002). 
 \bibitem{MOPAC} J.J.P.J. Stewart, Comput. Chem. {\bf 10}, 209, (1989). J.J.P.J. Stewart., Quant. Chem. Prog. Exch., {\bf 10}, 86, (1990).
% \bibitem{expreaction} Citare esperimento della reazione
 \bibitem{libro} W. Koch  and M.C. Holthausen, "A Chemist's Guide to Density Functional Theory" (2nd Ed.), Wiley-VCH, Weinheim, 2000. 
 \bibitem{laio} A. Laio and M. Parrinello, Proc. Nat. Acad. Sci. {\bf 99}, 12562, (2002)
 \end{thebibliography}
\end{document}